\documentclass[submission,copyright,creativecommons]{eptcs}
\providecommand{\event}{FROM 2025} 

\usepackage{iftex}

\ifpdf
  \usepackage{underscore}         
  \usepackage[T1]{fontenc}        
\else
  \usepackage{breakurl}           
\fi

\usepackage[dvipsnames]{xcolor}
\usepackage{amsmath}
\usepackage{amsthm}
\usepackage{cleveref}
\usepackage{graphicx}
\usepackage{subcaption}
\usepackage{wrapfig}
\usepackage{mathpartir}

\usepackage{calc}

\newcommand*{\maskM}[2]{%
  \mathord{\makebox[\widthof{\(#1\)}]{\(#2\)}}%
}

\usepackage{listings}
\newcommand{\Rocqinline}{\lstinline}
\lstnewenvironment{RocqCode}{}{}
\theoremstyle{definition}

\DeclareMathOperator{\sconj}{\ast\ast}
\DeclareMathOperator{\simpl}{-\ast}

\usepackage{tikz}
\usetikzlibrary{calc,positioning,arrows.meta,quotes,bending}

\title{Pleasant Imperative Program Proofs with GallinaC}
\author{
  Frédéric Fort\textsuperscript{1}
  \and
  David Nowak\textsuperscript{1}
  \and
  Vlad Rusu\textsuperscript{2}
  \and
  \institute{\textsuperscript{1}~Univ. Lille, CNRS, Centrale Lille, UMR 9189 CRIStAL, F-59000 Lille, France}
  \institute{\textsuperscript{2}~Univ. Lille, Inria, CNRS, Centrale Lille, UMR 9189 CRIStAL, F-59000 Lille, France}
}
\def\titlerunning{Pleasant imperative program proofs with GallinaC}
\def\authorrunning{F. Fort, D. Nowak, V. Rusu}

\begin{document}

\maketitle

\section{Introduction}
\label{sec:intro}

Even with the increase of popularity of functional programming,
imperative programming remains a key programming paradigm, especially
for programs operating at lower levels of abstraction. When such
software offers key components of a Trusted Computing Base (TCB),
e.g. an operating system kernel, it becomes desirable to provide
mathematical correctness proofs.

However, current real-world imperative programming languages possess
``expressive'', i.e. overly permissive, semantics. Thus, producing
correctness proofs of such programs becomes tedious and error-prone,
requiring to take care of numerous ``administrative''
details. Ideally, a proof-oriented imperative language should feature
well-behaved semantics while allowing typical imperative idioms.

To obtain a high-degree of confidence in the correctness of such a
language, its tools should be developed inside a proof-assistant such
that program proofs, be it that the compiler passes preserve semantics
or any proof written by language users, are machine checked.

Such an \emph{embedding} of a programming language may either be deep
or shallow~\cite{wildmoser2004}. With \emph{deep embedding}, the language is defined using
the available data structures of the implementation language,
e.g. inductive types in the Rocq proof
assistant~\cite{coq-rocq-refman}. With \emph{shallow embedding}, the
programming language is directly a subset of the implementation
language.

Historically, deep embedding has been considered the preferred
approach for sufficiently complex languages. Indeed, deep embedding
allows to define arbitrary complex languages independent from the
limitations of implementation language. For instance, Gallina, the
functional programming language of the Rocq proof assistant is total
for soundness reasons, excluding turing-completeness. Yet, inductive
proposition may define the semantics of a turing-complete language,
such as C~\cite{leroy2016}.

However our experience has shown that, deep embedding poses
practicability issues. Indeed, since the language is completely
constructed bottom-up, every property, theorem, proof tactic has to be
defined and proven from the ground up. On the opposite, since shallow
embedding uses a subset of the implementation language, if the
language is sound, certain theorems are for free and the usual proof
tactics are available.

Previous work using shallow embedding has produced substantial
programs and their correctness proofs such as a hypervisor with
formally-proven isolation
guarantees~\cite{jomaa2018formal,jomaa2018proof} or a real-time
scheduler using the Earliest-Deadline First
policy~\cite{vanhems2022}. However, their source languages were not
turing-complete. In particular, they lacked a generic recursion
mechanism: Their \Rocqinline{while} loop construct used fuel,
i.e. an at compile time defined maximum iteration count.

Recent work~\cite{cheval2024} has shown how to define partial
(co-)recursive functions directly inside the Gallina language. This
enables us to define a shallow embedding of a potentially unbounded
\Rocqinline{while} loop, bringing turing-completeness to the
Gallina language.

\subsection*{Contributions}
\label{sec:intro/contribution}

In this paper, we present \emph{GallinaC}, a shallow embedding of a
Turing-complete imperative language directly inside the functional
programming language of the Rocq proof assistant, Gallina. In
particular, it features a truly generic and unbounded
\Rocqinline{while} loop. Having a functional core means proofs
about GallinaC programs may use the same tactics as proofs about pure
functional ones.

Compilation from GallinaC to binary is possible through the CompCert
certified compiler. A chain of forward simulations guarantees that
compilation passes preserve semantics.

Work on GallinaC is still under progress, but we present first
promising results. A prototype implementation has shown the viability
of GallinaC with the correctness proof of a list reversal procedure
for linked-lists of unknown size. We currently focus on the forward
simulation between the GallinaC intermediate representation (IR) and
Cminor, the entry language of the CompCert back-end. The GallinaC
sources are available online (\url{https://gitlab.cristal.univ-lille.fr/ffort/gallinac}).

\section{Motivational Example}
\label{sec:running}

\begin{wrapfigure}{l}{0.5\linewidth}
  \begin{RocqCode}
Definition reverse ptr :=
  var node <- ptr;
  var new_next <- NULL;
  let deref_next :=
    do ptr <- read_var node;
    do val <- read_ptr (ptr + 1);
    ret val
  in
  let cond :=
    do curr <- read_var node;
    ret (negb (curr =? NULL))
  in
  while cond {{
    do curr <- read_var node;
    do next <- deref_next;
    do prev <- read_var new_next;
    write_ptr (ptr + 1) prev;;
    write_var node next;;
    write_var new_next curr
  }};;
  read_var new_next
  \end{RocqCode}

  \caption{List reversal in GallinaC}
  \label{fig:programs}
\end{wrapfigure}

As a motivational example, let us consider the program from
\Cref{fig:programs}. Provided a pointer that points to a member of a
simply-linked list, it performs an in-place list reversal by swapping
pointers. Linked list nodes are composed of two memory words, one
holding the value and one holding the pointer to the next list node.

The first two lines declare mutable variables which hold respectively
the list node currently under consideration and the node's new
\Rocqinline{next} pointer. Obviously, at program start these are
initialized to the list's entry pointer and the typical sentinel value
\Rocqinline{NULL}.

Next, two helper functions are declared. The first,
\Rocqinline{deref_next}, loads the list node pointed to by the current
node's \Rocqinline{next} pointer. The second, \Rocqinline{cond}, is
the while loop's invariant: the pointer to the next node to consider
is not \Rocqinline{NULL}.

The remainder is the ``main'' function of the program. The core is a
\Rocqinline{while} loop that breaks once \Rocqinline{node} holds the
\Rocqinline{NULL} pointer. At each loop iteration, the current node's
\Rocqinline{next} field is updated. For the next iteration,
\Rocqinline{node} and \Rocqinline{new_next} are set respectively to
the current node's \Rocqinline{next} field and the current node. The
program terminates by returning the address of the list's new first
node.

This program illustrates GallinaC's new \Rocqinline{while} loop. Note
that the loop is not annotated with a fuel amount or any other kind
of restriction: a programming error could produce an infinite loop.

\section{Language Design}
\label{sec:lang}

GallinaC programs are Gallina terms with type
\Rocqinline{program S A}.
The \emph{program monad} incorporates state, failure and
non-termination effects. While it could theoretically be constructed
using monad transformers, its current implementation is a custom
monad. The type parameter \Rocqinline{S} is the type of the state,
i.e. the heap and store of the program. The type parameter
\Rocqinline{A} is the return type of the monadic computation, assuming
it doesn't fail.

The non-termination monad is adapted from~\cite{cheval2024}. The monad
functor is the \Rocqinline{option} type where \Rocqinline{None}
encodes bottom, i.e. non-termination. We first define the
\Rocqinline{while} \emph{functional}:

\begin{center}
  \begin{RocqCode}
    Definition whileF
    {S : Type} (cond : program S bool)
    (W : program S unit -> program S unit)
    (body : program S unit) : program S unit :=
    If cond then (body ;; W body) else ret tt.
  \end{RocqCode}
\end{center}

As can be seen, the recursive, non-guarded, call is replaced by a call
to an additional argument \Rocqinline{W}. We then
leverage Kleene's fixed-point theorem stating that a continuous
function with signature $A \rightarrow A$ where A is equipped with a
CPO has a least fixed-point. For our \Rocqinline{whileF} this
fixed-point is the ``true'' \Rocqinline{while}. Continuity itself is
proven using Haddock's theorem~\cite{cheval2024}.

\begin{figure}
  \centering
  \begin{subfigure}[t]{0.45\linewidth}
    \centering
    \begin{align*}
      e & = id \\
        & \maskM{ = \hspace{1em}}{\hspace{1ex} | \quad} tt \\
        & \maskM{ = \hspace{1em}}{\hspace{1ex} | \quad} true\ |\ false\ |\ ! e\ |\ ... \\
        & \maskM{ = \hspace{1em}}{\hspace{1ex} | \quad} 0\ |\ 1\ |\ 2\ |\ ... \\
        & \maskM{ = \hspace{1em}}{\hspace{1ex} | \quad} NULL \\
        & \maskM{ = \hspace{1em}}{\hspace{1ex} | \quad} e + e\ |\ e = e\ |\ ...
    \end{align*}
    \caption{GallinaC expressions}
    \label{fig:exprcmd/expr}
  \end{subfigure}
  \quad
  \begin{subfigure}[t]{0.45\linewidth}
    \centering
    \begin{align*}
      cmd & = \hspace{1ex} \mathtt{ret}\ e \\
          & \maskM{ = \hspace{1em}}{\hspace{1ex} | \quad} \mathtt{do}\ x\ \texttt{<-}\ cmd\ \mathtt{;}\ cmd \\
          & \maskM{ = \hspace{1em}}{\hspace{1ex} | \quad} cmd\ \mathtt{;;}\ cmd \\
          & \maskM{ = \hspace{1em}}{\hspace{1ex} | \quad} id (e, e, ...) \\
          & \maskM{ = \hspace{1em}}{\hspace{1ex} | \quad} \mathtt{if}\ e\ \mathtt{then}\ cmd\ \mathtt{else}\ cmd \\
          & \maskM{ = \hspace{1em}}{\hspace{1ex} | \quad} \mathtt{while}\ cmd\ \mathtt{\{\{}\ cmd\ \mathtt{\}\}} \\
          & \maskM{ = \hspace{1em}}{\hspace{1ex} | \quad} \mathtt{var}\ v\ \texttt{<-}\ cmd\ \mathtt{;}\ cmd \\
          & \maskM{ = \hspace{1em}}{\hspace{1ex} | \quad} \mathtt{read\_var}\ v \\
          & \maskM{ = \hspace{1em}}{\hspace{1ex} | \quad} \mathtt{write\_var}\ v \ e\\
          & \maskM{ = \hspace{1em}}{\hspace{1ex} | \quad} \mathtt{alloc}\ n\ e \\
          & \maskM{ = \hspace{1em}}{\hspace{1ex} | \quad} \mathtt{read\_ptr}\ e \\
          & \maskM{ = \hspace{1em}}{\hspace{1ex} | \quad} \mathtt{write\_ptr}\ e\ e \\
          & \maskM{ = \hspace{1em}}{\hspace{1ex} | \quad} \mathtt{free}\ e
    \end{align*}
    \caption{GallinaC commands}
    \label{fig:exprcmd/cmd}
  \end{subfigure}

  \caption{GallinaC expressions and commands}
  \label{fig:exprcmd}
\end{figure}

For convenience, we offer the classical monadic notations to simplify
the writing of GallinaC programs: return (\Rocqinline{ret}) and bind
(\Rocqinline{do x <- ret 5; ret (x + 3)}), as well as immediate
failure (\Rocqinline{fail}), infinite loop (\Rocqinline{loop}) and
discarding bind (\Rocqinline{some_side_effect ;; ret true}).

The language distinguishes \emph{expressions} from \emph{commands}.
Expressions include operations on variables, the unit expression
\Rocqinline{tt}, booleans, and natural numbers with fixed-width
unsigned integer semantics, as well as pointers.

Commands contain the side-effectful part of the language. Monadic
return and bind use the typical notations and semantics. Control-flow
(function calls, \texttt{if} branches and turing-complete
\texttt{while} loops) is reserved to commands. Note in particular that
the condition of the \texttt{while} loop is itself a command.

A \emph{store} is accessible using the \texttt{var}-family of
commands. Variables in the store are accessible by their name and
possess a unique location in memory for their entire lifetime,
enabling both reads and writes to them. This distinguishes them from
variables bound with \texttt{do} which may only be read from once set
and may be moved around and dropped by the compiler for optimization
reasons. A similar distinction exists in CompCert between temporaries
and locals which are respectively equivalent to \texttt{do} and
\texttt{var} variables.

A \emph{heap} is accessible using the \texttt{alloc}, \texttt{free}
and \texttt{*\_ptr} commands. Pointers are distinct from integers in
that they possess not only an address but also a
\emph{provenance}~\cite{memarian2019}. A provenance records the memory
region a pointer may access, typically the region it was allocated
from. It is a programming error to use a pointer to access memory
outside its provenance. Note that this does not prevent to use a
pointer to a data structure field to access other fields, nor to
traverse a linked-list by following successive pointers. Provenance
formalizes concepts already existing in the C
standard~\cite{wg14-dr260,isoc2011} and MISRA C~\cite{bagnara2018}.

\section{Shallow Separation Logic}
\label{sec:seplog}

\begin{figure}
  \centering


  \begin{RocqCode}
Definition Pred := S -> Prop.

Definition star (P R: Pred): Pred :=
  fun s => exists s1 s2,
    store s = store s1 /\ store s = store s2 /\
    (* (heap s) can be partitioned into (heap s1) (heap s2) *)
    Partition (heap s) (heap s1) (heap s2) /\
    P s1 /\ Q s2.
Infix "**" := star.

Definition wand (P R: Pred): Pred :=
  fun s => forall s' hp,
    hp $= heap s $++ heap s' -> store s' = store s ->
    P s' -> R (mkState (store s') hp).
Infix "-*" := wand.

  \end{RocqCode}

  \caption{Shallow separation logic definitions}
  \label{fig:seplog/shallow}
\end{figure}

Separation logic~\cite{ohearn2019,reynolds2002,reynolds2005} is an
extension of Hoare logic~\cite{hoare1969} aimed at reasoning about
mutable data structures with pointers. Recall that the %
\emph{Hoare triple} %
$\{\ P\ \}\ c\ \{\ Q\ \}$ states that executing some code $c$ in a state
where proposition $P$ holds does not produce a ``crash'' (more
specifically any incoherent state) and that proposition
$Q$ holds in the final state.

Separation logic introduces two logical connectives, $\sconj$ and
$\simpl$, called respectively \emph{separating conjunction} and
\emph{separating implication}. If the separating conjunction
$P \sconj Q$ holds, then the program state can be divided into two
clearly disjoint parts such that $P$ holds for the first part and $Q$
for the second part. The separating implication $P \simpl Q$ states
that if the current heap were extended such that $P$ held, then $Q$
would also hold. More intuitively, if code execution resulted in a
state where $P$ held, then so would $Q$.

Together with the \emph{frame rule}, separation logic simplifies the
proof process for realistic imperative programs. It states that any
Hoare triple $\{\ P\ \}\ c\ \{\ Q\ \}$ can be extended to a more
complete context $\{\ P \sconj R\ \}\ c\ \{\ Q \sconj R\ \}$ provided
that $c$ does not mutate any free variables in $R$.

Logically, our implementation of separation logic is shallow as well.
We solved common issues with shallow embeddings~\cite{wildmoser2004}
by leveraging Rocq's automated proof-search utilities. Lemmas may be
collected in proof databases which the \Rocqinline{auto/eauto} tactics may
leverage to solve the given goal.

Moreover, note that unlike~\cite{nanevski2008}, we opted for an
``initially weakly-typed'' approach to separation logic. All programs
have a type of the form \Rocqinline{program S A}. Predicates are a proof-only
construct added on top. Indeed, in our experience working both with
system and proof engineers, requiring the right proof annotations
upfront does not scale to large programs such as OS
kernels~\cite{jomaa2018formal}.

\Cref{fig:seplog/shallow} shows the definitions of the core
definitions. Predicates $P, R$ discussed above are terms with type
$S \rightarrow Prop$, i.e. propositions which depend of the
state. Output predicates like $Q$ above additionally may depend upon
the output produced by the program code.

\emph{Separating conjunction} is defined as \Rocqinline{star} (a notation for
the infix \Rocqinline{**} is provided), a predicate stating that for
any state \Rocqinline{s}, we can define sub-states \Rocqinline{s1} and
\Rocqinline{s2} such that their stores are equal to the store of
\Rocqinline{s}, their heaps are a partitioning of the heap of
\Rocqinline{s}, and \Rocqinline{P} and \Rocqinline{R} hold in their
relative states.
\emph{Separating implication} is defined as predicate \Rocqinline{wand} (and a notation
\Rocqinline{-*}) stating that for any states \Rocqinline{s},
\Rocqinline{s'} and heap \Rocqinline{hp}, if \Rocqinline{hp} is the
union of the states' heaps and \Rocqinline{P} is true in
\Rocqinline{s'}, then \Rocqinline{R} is true in the state with the
extended heap.

\section{Architecture}
\label{sec:archi}

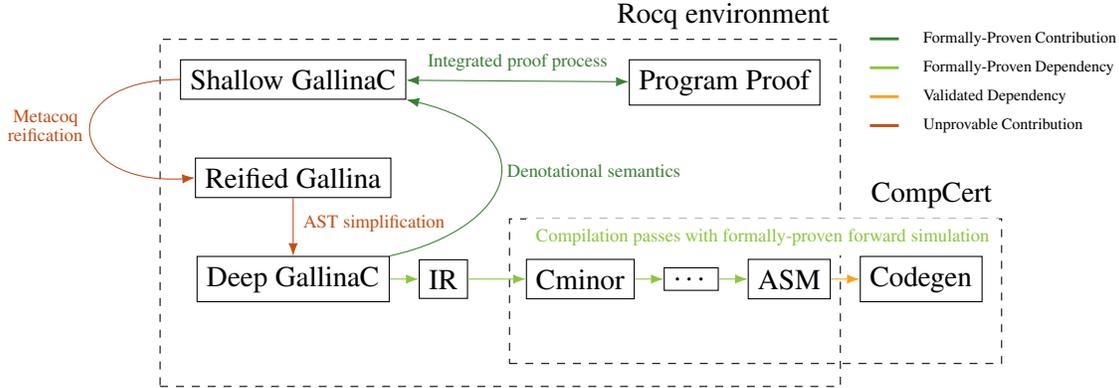
\begin{figure}
  \centering
  \begin{tikzpicture}[x=1em,y=1em,fwsim /.tip= . Latex]

    \node[draw, dashed, minimum width=23.5em, minimum height=12em] (Rocqbox) {};
    \node[above left=0 of Rocqbox.north east] (Rocqname) {Rocq environment};

    \node[draw, below right=1em of Rocqbox.north west] (ShallowGallinaC) {Shallow GallinaC};


    \node[draw, below left=1em of Rocqbox.north east] (GallinaCProof) {Program Proof};

    \node[draw, below=2em of ShallowGallinaC] (ReifiedGallina) {Reified Gallina};

    \node[draw, below=2em of ReifiedGallina] (DeepGallinaC) {Deep GallinaC};

    \node[draw, right=1em of DeepGallinaC] (IR) {IR};

    \node[draw, dashed, above right=2em of IR.north east, anchor=north west, minimum width=17em, minimum height=5 em] (CompCertbox) {};
    \node[above left=0 of CompCertbox.north east] (CompCertname) {CompCert};

    \node[draw, right=2em of IR] (Cminor) {Cminor};
    \node[draw, right=1em of Cminor] (CCdots) {\ldots};
    \node[draw, right=1em of CCdots] (ASM) {ASM};
    \node[draw, right=1em of ASM] (CCCodegen) {Codegen};

    \draw [ fwsim-fwsim,OliveGreen] (ShallowGallinaC) edge ["\fontsize{7pt}{7pt}\selectfont{}Integrated proof process"] (GallinaCProof);


    \draw [ -fwsim,Bittersweet] (ShallowGallinaC.west) .. controls +(-4.5,0) and +(-4.5,-0.5) .. (ReifiedGallina.west);
    \node [left=3em of ShallowGallinaC.south west, anchor=north east, Bittersweet,align=center,font=\fontsize{7pt}{7pt}\selectfont] (MetaCoqReification) {Metacoq\\reification};

    \draw [-fwsim,Bittersweet] (ReifiedGallina) -- (DeepGallinaC);
    \node[below=0.25em of ReifiedGallina.south,anchor=north west,Bittersweet,align=center,font=\fontsize{7pt}{7pt}\selectfont] (ASTSimplification) {AST simplification};

    \draw [-fwsim,LimeGreen] (DeepGallinaC) -- (IR);
    \draw [-fwsim,LimeGreen] (IR) -- (Cminor);
    \draw [-fwsim,LimeGreen] (Cminor) -- (CCdots);
    \draw [-fwsim,LimeGreen] (CCdots) -- (ASM);
    \draw [-fwsim,YellowOrange] (ASM) -- (CCCodegen);

    \node[above=0.75em of Cminor.north west,anchor=west, LimeGreen, fill=white, align=center,font=\fontsize{7pt}{7pt}\selectfont] (ForwardSimulation) {Compilation passes with formally-proven forward simulation};

    \draw [-fwsim,OliveGreen] (DeepGallinaC.north east) .. controls +(4.5em,1em) and +(5em,-1em) .. (ShallowGallinaC.south east);
    \node[right=3.65em of ReifiedGallina.north east, anchor=west, yshift=-5pt, OliveGreen, align=center, font=\fontsize{7pt}{7pt}\selectfont] (Denotation) {Denotational semantics};

    \draw [OliveGreen, line width=0.1em] ($(Rocqbox.north east) + (1em, 0)$) -- ($(Rocqbox.north east) + (2em, 0)$);
    \node [anchor=west] (LegendOurProven) at ($(Rocqbox.north east) + (2.5em, 0) $) {\tiny Formally-Proven Contribution};
    \draw [LimeGreen, line width=0.1em] ($(Rocqbox.north east) + (1em, -1em)$) -- ($(Rocqbox.north east) + (2em, -1em)$);
    \node [anchor=west] (LegendExtProven) at ($(Rocqbox.north east) + (2.5em, -1em) $) {\tiny Formally-Proven Dependency};
    \draw [YellowOrange, line width=0.1em] ($(Rocqbox.north east) + (1em, -2em)$) -- ($(Rocqbox.north east) + (2em, -2em)$);
    \node [anchor=west] (LegendExtProven) at ($(Rocqbox.north east) + (2.5em, -2em) $) {\tiny Validated Dependency};
    \draw [Bittersweet, line width=0.1em] ($(Rocqbox.north east) + (1em, -3em)$) -- ($(Rocqbox.north east) + (2em, -3em)$);
    \node [anchor=west] (LegendExtProven) at ($(Rocqbox.north east) + (2.5em, -3em) $) {\tiny Unprovable Contribution};

  \end{tikzpicture}
  \caption{GallinaC workflow}
  \label{fig:workflow}
\end{figure}

The typical workflow when using GallinaC is presented in
\Cref{fig:workflow}. \textcolor{OliveGreen}{Green} arrows represent
computation steps which are associated with formal
proofs. \textcolor{YellowOrange}{Orange} arrows represent computation
of not formally-proven dependencies, but for which validation tools
exist. \textcolor{Bittersweet}{Red} arrows represent computation steps
which are essentially unprovable, we will discuss how to address this
below. Recall that GallinaC programs are monadic Gallina terms with
type \Rocqinline{program S A} where \Rocqinline{S} is the heap and
store and \Rocqinline{A} is the return type.

Development starts inside the ``perfect Rocq world'' where programs
can be composed and defined using typical Rocq commands
(\Cref{fig:programs}). From there, propositions on these programs can
be proven directly on the monadic program using the typical commands
and tactics for proofs. To produce a binary, an extraction process
operates on a deep embedding (\Cref{fig:shallowdeep}). A chain of
forward simulation proofs guarantees the compilation process'
correctness.

\subsection{Reification}

Producing a binary which may be used outside the Rocq proof assistant
requires an \emph{extraction plugin}. However, the default extraction
plugins target high-level garbage-collected functional languages
(OCaml, Haskell) which are incompatible with bare-metal
programming. Thus, we provide a custom extraction path relying on
MetaCoq~\cite{sozeau2020} and CompCert~\cite{leroy2016}.

Provided a Rocq term \Rocqinline{t}, MetaCoq allows to \emph{reify}
it into an inductive term representing the term's AST. This inductive
term is then processed, resulting in a AST for deeply-embedded
GallinaC as shown in \Cref{fig:shallowdeep/deep}.

\begin{figure}
  \centering

  \begin{subfigure}[t]{0.45\linewidth}
    \begin{RocqCode}
Definition deref_next :=
  do ptr <- read_var node;
  do val <- read_ptr (ptr + 1);
  ret val.
    \end{RocqCode}
    \caption{Shallow GallinaC}
    \label{fig:shallowdeep/shallow}
  \end{subfigure}
  \quad
  \begin{subfigure}[t]{0.45\linewidth}
    \begin{RocqCode}
Definition deref_next_deep :=
  c_bind "ptr" (c_read_var node)
    (c_bind
      "val"
      (c_read_ptr
        (e_ptr_shift_fw
          (e_fvar "ptr")
          (e_nat 1)))
      (c_ret (e_fvar "val"))).
    \end{RocqCode}
    \caption{Deep GallinaC}
    \label{fig:shallowdeep/deep}
  \end{subfigure}
  \quad
  \begin{subfigure}[t]{0.45\linewidth}
    \begin{RocqCode}
Lemma reverse_eq:
  denote deref_next_deep = deref_next.
Proof.
  reflexivity.
Qed.
    \end{RocqCode}
    \caption{Correctness by denotation}
    \label{fig:shallowdeep/denote}
  \end{subfigure}

  \caption{Shallow and Deep GallinaC comparison}
  \label{fig:shallowdeep}
\end{figure}

Note however, that this reification process is happening necessarily
outside the Rocq proof assistant, producing a break in trust. To
regain confidence in the deeply-embedded GallinaC, we equip it with
a denotational semantics.

The denotational semantics undoes the reification process directly
inside the proof assistant. Thus, if the reification process is
correct, denotation should yield the original shallow GallinaC
program. In essence, while we can't prove the reification process
correct for \emph{all} programs, we can show it correct for \emph{any}
program by simple reflexivity~(\Cref{fig:shallowdeep/denote}).

\subsection{Compilation}


Instead of writing our own compiler, we rely on
CompCert~\cite{leroy2016}, a formally-verified compiler for a
reasonable subset of C. Our entry-point is the Cminor language, the
highest-level language of the back-end~\cite{leroy2009}. While still
resembling C, Cminor is much ``simpler'', e.g. types are completely
discarded and the stack is a primitive value that must be allocated
with the correct size at function entry. To bridge the gap between the
deep GallinaC and Cminor, we first compile to an Intermediate
Representation (IR).

Our IR still resembles GallinaC, with a distinction between commands
and expressions for instance, but it simultaneously follows the
conventions of CompCert languages. For instance, identifiers are no
more strings, but positives. Moreover, all side-effectful operations
are both performed in lockstep inside the CompCert memory and the
GallinaC state to guarantee coherence.

Once the IR is translated to Cminor, the usual CompCert passes are
applied. Note in particular that we did not patch CompCert, we use for
all our developments the publicly available stable version. The
correctness of these compilation passes is guaranteed by a chain of
forward simulations, i.e. the proof that the behaviors of the source
program are preserved in the compiled program.

Note that while we verified the correctness of the MetaCoq reification
using a denotational semantics, CompCert uses for its compilation an
operational semantics. As such, the step from deep GallinaC to our IR
could potentially distort the semantics of our programs. To dispel all
such doubts, we have proven that the denotational and operational
semantics of deep GallinaC \emph{agree}. More specifically:
evaluation of the operational semantics preserves denotation.

Finally, the last step, code generation, must necessarily execute outside the
Rocq environment. This last step is thus, necessarily less
trusted. The CompCert team having been faced with the same issues,
they provide tools to validate such programs
nonetheless~\cite{kastner2018,leroy2016}.

\section{Prototype}
\label{sec:results}

Work on GallinaC is still ongoing. We developed a first prototype
version to show the feasibility of our approach. We programmed the
list reversal program shown above, showed its correctness and
extracted a binary from it. However, many aspects were left explicitly
simplistic. Most notably, the store consisted only of two global
read-write variables and there existed no distinction between
integers, pointers and addresses.

This approach however hit a wall when trying to prove the forward
simulation of compilation passes. Thus, a considerable amount of time
was invested in designing sound interfaces for the different aspects
of imperative programming. These new interfaces introduced breaking
changes in the user-facing monadic language. Our current approach is
bottom-up and we are focusing on the proof of forward simulation
between the IR and Cminor. We will return to the user-facing language
and tooling after this.

We plan also to increase the expressivity of the language. For
instance, early loop exit is not yet supported. To preserve the
adequacy, with program proofs, we are looking into replacing the
program monad with a Freer or Continuation monad.

\section{Conclusion and Future Work}
\label{sec:conclusion}

In this paper, we presented GallinaC, a collection of programming
tools for formally-proven low-level imperative code. Developers write
programs in a monadic subset of Rocq's functional language,
Gallina. Proofs about these programs may be performed directly in the
same programming environment and separation logic makes proofs
feasible. The extraction to concrete binary code is performed in two
steps. First, MetaCoq allows to reify the program into a deep
embedding. A denotational semantics guarantees the correctness of this
step. Second, the deep embedding is compiled using the CompCert
compiler to assembly. A chain of forward simulation proofs on
operational semantics guarantee the correctness of this process. We
have proven the equivalence of these semantics, showing that the
compilation process is truly correct end-to-end.

Current work focuses on integrating all components and finalizing the
correctness proofs of compilation. In future work, we will want to
design tools to guarantee that \emph{any} proof on monadic GallinaC
holds for the compiled assembly.

\bibliographystyle{eptcs}
\bibliography{ffort-bib/compil,ffort-bib/embedded,ffort-bib/software,ffort-bib/pip,ffort-bib/proof,ffort-bib/standards}

\end{document}